\documentclass[prl,showpacs,amsfonts,twocolumn,superscriptaddress]{revtex4-1}
\usepackage{amsmath,amssymb,graphicx,hyperref, color}
\renewcommand{\vec}[1]{\mathbf{#1}}

\newcommand{\gss}{\text{eff}} 
\newcommand{\Oz}{\Omega_{\perp}}
\newcommand{\Oxy}{\Omega_{\parallel}}

\newcommand{\ls}{\ell_0}

\begin{document}

\title{Controllable half-vortex lattices in an incoherently pumped polariton condensate}
\author{Jonathan Keeling} 
\affiliation{SUPA, School of Physics and Astronomy, University of St Andrews, 
  KY16 9SS, UK}
\author{Natalia G. Berloff}
\affiliation{Department of Applied Mathematics and Theoretical Physics,
  University of Cambridge, CB3 0WA, UK}

\date{\today}
\begin{abstract}
  We show how the transition between synchronized and desynchronized
  states of a spinor polariton condensate can be used to drive a
  transition between stationary vortex lattices and half-vortex
  lattices.  This provides a way to control polariton spin textures by
  a combination of pump spot profile and applied magnetic fields. To
  do this, we extend the model of non-equilibrium spinor condensates
  to include relaxation, and study how this affects the
  desynchronization transition. We discuss how the pattern formation
  can be explained by behavior of the homogeneous system.
\end{abstract}
\maketitle

Among the various topological defects that may exist in superfluids
and superconductors, quantized vortices receive particular theoretical
and experimental attention.  These localized nonlinear structures are
nontrivial low energy excitations of the system, and so knowledge of
the core structure, topology, symmetry, dynamics and interactions of
vortices reveals details of the low energy properties of the system.
As well as conventional singular quantized vortices seen in
${}^4$He-II superfluids, one component Bose-Einstein condensates, and
s-wave superconductors, there also exist more exotic quantized
vortices, such as non-singular vortices, continuous vortices and
composite vortices~\cite{volovik03}.  Non-singular vortices have a
hard core set by the coherence length but with a non-zero order
parameter within the core.  Such order parameter textures have been
explored in various systems, including multi-component cold atomic
gases~\cite{Ruben10}, nonlinear optics~\cite{Pismen94}, superfluid
${}^3$He-A\cite{volovik03} and
Sr$_2$RuO$_4$\cite{mackenzie03}. However, the control of these
structures and, in particular, their nucleation from conventional
singular vortices has been elusive. In this letter we propose how
these structures can be controllably created and manipulated in
polariton condensates by using magnetic fields.  We make use of the
non-equilibrium properties of the polariton condensate, and so
illustrate how a nonequilibrium condensate may show behavior not
present in equilibrium spinor condensates like cold atoms or
${}^3$He-A.

Microcavity polaritons are quasiparticles that result from the
hybridization of excitons and photons confined inside semiconductor
microcavities. At low temperatures, they can condense, and polariton
condensation has been the subject of much recent work
(see~\cite{deng10} for a recent review).  Polariton condensates differ
from condensates of ultracold atomic gases or helium in that polariton
condensates can be non-equilibrium systems: the photon component
decays, and so new polaritons must be injected into the system to
maintain a steady state.  Hence, polariton condensates can be thought
of as sustained nonequilibrium systems with nontrivial spatiotemporal
properties.  Spatiotemporal pattern formation has been extensively
studied in systems such as fluid and plasma dynamics, reaction
diffusion systems, biological structures, and neural networks.  Order
parameter equations describe the structure and dynamics of pattern
formation in these systems, and the order parameter equation for the
polariton system (and closely related nonlinear optical
systems~\cite{staliunas03}) is the complex Ginzburg-Landau
equation (cGLE) ~\cite{aranson02}.  Polariton condensates are well
suited for studying pattern formation since  the source term can be
made to vary in space (and time) and thereby control pattern
formation.  Recent experiments on polaritons have begun to explore
this possibility, by studying phenomena such as quantized
vortices~\cite{lagoudakis08}, soliton propagation~\cite{amo09:bullet},
and patterns in one-dimensional samples~\cite{wertz10}.

As polaritons have a polarization degree of freedom, they can support
unconventional quantized vortices, such as non-singular vortices where
only one polarization component has
vortices~\cite{rubo07}.  Such polariton
``half-vortices'' have been seen in experiments~\cite{lagoudakis09},
but there are questions about the conditions required for them to
remain stable and
independent~\cite{Flayac10,*solano10-2}. Experimental investigation of
their stability is hard because observations of polariton vortices
have generally relied on disorder to pin
vortices~\cite{lagoudakis08,lagoudakis09}, and half-vortices could be
stabilized by disorder induced strain splitting of polarization
states.  Theoretical alternatives to these disorder-localized vortices
have been proposed, such as rotating lattices in harmonically trapped
incoherently pumped condensates~\cite{keeling08:gpe}.  Direct
observation of such rotating vortex lattices is
however hard with current approaches to observing
vortices~\cite{roumpos10,lagoudakis11}.  There are also proposals to
create half vortices using parametric~\cite{gorbach10,pigeon11} or
coherent pumping~\cite{Liew08}.  None of these schemes combine the
necessary features to experimentally investigate the stability of
half vortices in a clean incoherently pumped system.  This is the task
we address in this letter, by showing how interplay between the
synchronization-desynchronization transition of the homogeneous
system~\cite{wouters08:prb,borgh10} and interference between
polaritons originating from separate incoherent pumping spots can
produce non-rotating lattices of vortices, where a transition between
separate and locked half-vortices can be controlled by an applied
magnetic field.

We start by introducing the cGLE for a spinor polariton condensate.
This includes a term describing pumping, and for stability this
must be nonlinear, so that it saturates as the condensate
density increases.  In a single component condensate, the simplest
such nonlinearity is to write, $i \hbar \partial_t \psi \sim i (\gamma
-\Gamma |\psi|^2) \psi$.  However, as discussed
recently~\cite{wouters10:superfluid,*wouters10,*liew10}, adding an energy dependent gain
term $i \hbar \partial_t \psi \sim i(\gamma - \Gamma |\psi|^2 - i\eta
\hbar \partial_t) \psi$ can be important to correctly describe
pattern formation where energy relaxation plays an important r\^ole.
The (dimensionless) parameter $\eta$ is the rate at which gain
decreases with increasing energy. In the spinor case, a further
possibility emerges:  there can be a cross-polarization term
whereby the gain of one component may  be affected by the density
of the other component.  These various nonlinearities can all be
understood in terms of transfer between an excitonic reservoir and the
condensate, driven by the lack of chemical
equilibrium~\cite{griffin09}. Nonlinearity arises both from depletion
of the reservoir, and from shifts of the condensate chemical
potential.  We have therefore included all three types of possible
nonlinearity, and studied how they affect the behavior.   The spinor cGLE
is thus:
\begin{multline}
  \label{eq:1}
   i \hbar \partial_t \psi_{\pm} = 
    \bigl[
    \pm \Oz -
    \frac{\hbar^2 \nabla^2}{2 m} 
    + U_0 |\psi_{\pm}|^2 + (U_0 - 2 U_1) |\psi_{\mp}|^2 
    \bigr.\\\bigl.
    +i P(r, \psi_\pm)  -i\kappa \bigr] \psi_{\pm} +
    \Oxy \psi_{\mp},
\end{multline}
where $\pm$ stand for left- and right- polarizations.  We first
discuss the parameters describing energetics, and then turn to those
related to the pump configuration and loss.  

Since the nonlinear terms can be rescaled by a change of density, the
only important parameter for the interactions is the ratio $U_1/U_0$
which we take as $U_1/U_0=0.55$ (see e.g.~\cite{borgh10}). This
implies a weak attraction between opposite polarizations of
polaritons.  The parameter $\Oz$ represents a magnetic field
perpendicular to the 2D polariton condensate, while $\Oxy$ represents
either an in-plane field, or the intrinsic splitting of linear
polarizations present in most microcavity samples.  We consider values
based on the recent experiment in GaAs\cite{Larionov10}, for which a
range $0<\Oz<50\mu$eV was achieved, resulting in a transition from
linear to circular polarization. The splitting $\Oxy$ was only
constrained to be less than the linewidth.  We choose a value
$\Oxy=3\mu$eV consistent with this, which, as discussed below,
reproduces results for the homogeneous system that may be compared to
experiment.  Finally, the polariton mass only controls the overall
length scale $\ls = \sqrt{\hbar^2/2 m \kappa}$; where necessary we
give lengths in terms of this scale which would be $\ls = 1.95 \mu$m
for $m = 10^{-4} m_e$, taking decay rate $\kappa=0.1$meV.

We propose to form vortex lattices by pumping in three localized spots
equidistant from each other.  This means that the pump term
$P(r,\psi_\pm)$ is taken to be zero outside the pump spots. Inside each
pump spot, including the nonlinearities discussed above, it has the
form $P(r, \psi_\pm) = \gamma - \eta i \hbar \partial_t - \Gamma_s
|\psi_{\pm}|^2 - \Gamma_x |\psi_{\mp}|^2 $.  We consider pumping at
$2.5$ times threshold, i.e. $\gamma=2.5\kappa$. The parameter $\eta$
is not well constrained, so we consider a range of values.  For the
other nonlinearities, we take $\Gamma_s/U_0=0.3, \Gamma_x/U_0=0.1$,
based on values discussed in~\cite{borgh10}. The term $\Gamma_x$,
describing a cross-spin nonlinearity is new, and we do not have an
independent estimate of its value.  However, as long as
$\Gamma_x<\Gamma_s$, we find that the effect of increasing
$\Gamma_x/\Gamma_s$ is similar to the effect of increasing $\eta$.  If
$\Gamma_x = \Gamma_s$, the pumping process becomes spin-blind, and so
the polarization state is fixed purely by the energetics.

\begin{figure}[htpb]
  \centering

  \includegraphics[width=3.2in]{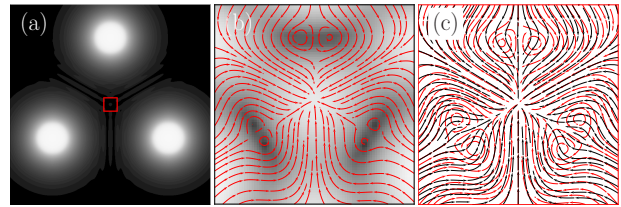}  
  \caption{Vortex lattices due to polariton flow from three pumping
    spots.  Panel (a) shows a $40 \ls \times 40 \ls $ region,
    illustrating the location of the pumping spots and a smaller $2\ls
    \times 2\ls$ region marked by a (red) square.  Panel (b) shows the
    streamlines and density profile of the majority component at
    $\Oz=2.4\mu$eV in the synchronized regime.  (The minority
    streamlines are locked and thus equivalent).  Panel (c) Show the
    streamlines for both polarizations (Red (gray)-- majority, Black
    -- minority) in the desynchronized regimes for $\Oz=100\mu$eV All
    figures are for $\gamma = 2.5 \kappa, \Oxy=3\mu\text{eV},
    \eta=0.1$.}
  \label{fig:vortices}
\end{figure}

If the pumping spots are irregularly spaced, then the generated
currents interfere and form vortex pairs which may lead to turbulent
state of interacting vortices as was discussed
in~\cite{turbulence10}. However if the spots are placed sufficiently
symmetrically then stationary vortex lattices form.  We performed
series of numerical experiments pumping at three spots of radius $R=5
\ls$ centered at $(0,20)\ls, (\pm 10\sqrt{3}, 10)\ls$ (see
Fig.~\ref{fig:vortices}). The numerical integration of Eq.~(\ref{eq:1})
was done by the fourth order finite differences in space and the
fourth order Runge-Kutta in time. Vortex lattices form as long as the
flux leaving one spot is enough to reach the adjacent spots, and thus
cause synchronization; this is very similar to what was found for
polariton condensates trapped in disorder
potentials~\cite{wouters08:prb}, however here we have no trapping
potential.  Vortex lattices are thus visible only for large enough
$\gamma/\kappa$ and small enough $\eta$.  Fig.~\ref{fig:vortices} is
for parameters where synchronization between spots just occurs, and as
a result, the lattice is not perfectly hexagonal; for larger $\alpha$,
lattices are more hexagonal (not shown).  Synchronization of pump
spots, and the consequent vortex lattices, can also survive some
degree of imperfection in size and location, 10\% variations can be
tolerated for the parameters used in Fig.~\ref{fig:vortices}.

For conditions where a vortex lattice occurs, we then vary the
detuning $\Oz$ and observe the evolution of vortex lattices.  As $\Oz$
increases, there is a transition from an elliptically polarized
condensate to a state where the two circular polarization components
desynchronize, producing two separate condensates, with two separate
chemical potentials (frequencies).  In the synchronized regime,
integer vortex lattices are formed with vortices of the same
circulation of the two polarizations sitting on the top of each other.
In the desynchronized regime, the lattice of the majority component
survives, but a number of possible behaviors for the minority
component can occur, depending on the values of $\gamma/\kappa, \eta,
\Oz$.  In Fig.~\ref{fig:vortices}(c), the minority component has a
vortex-free state.  There can also be cases where a vortex
lattice survives for the minority component, but because the
polarization components are not synchronized, the position of these
minority vortices can oscillate in time.  For large
$\gamma/\kappa$, these oscillations are small, and only a slight shift
of the minority lattice is visible.  However, for $\gamma = 2.5
\kappa$ as shown in Fig.~\ref{fig:vortices}(c), there is a clear
transition to a half-vortex lattice.  At much larger $\Oz$, a
transition back to a synchronized state is seen, but for $\eta=0.1$,
this is far beyond the range of $\Oz$ achievable in experiment.

Having illustrated how the desynchronization transition has dramatic
effects on the vortex lattices, we derive analytic approximations for
the period of the vortex lattice, and the critical field for the
desynchronization transition, by using a generalized local-density
approximation (considering also the phase gradients), and making use
of results for a homogeneously pumped system.  The period of the
vortex lattice can be understood by finding the wavelength of the
phase variation at points far from a pump spot.  If the pump spot is
large then currents can be neglected within the pump spot, and the
homogeneous system used to predict whether the system is synchronized,
and the values of chemical potential.  When synchronized,
$i \partial_t \psi_\pm =\mu \psi_\pm$, and so Eq.~(\ref{eq:1}) becomes
a pair of time-independent differential equations.  To solve these, it
is convenient to write $\psi_{\pm} = \sqrt{\rho_{\pm}} e^{i(\phi \pm
  \theta/2)}$.  In the synchronized case $\theta$ is constant so the
lattice spacing is set by the velocity $\vec{u} = \hbar \nabla
\phi/m$.  Far away from the pump spot, the pump term vanishes, and the
density decays and so interaction terms $U \rho_{\pm}$ can be
neglected. With these approximations the real and imaginary parts of
Eq.~(\ref{eq:1}) become:
\begin{math}
  \mu - \frac{1}{2}m|\vec{u}|^2 \mp \Oz = 
  \Oxy \sqrt{ {\rho_\pm}/{\rho_\mp} } \cos(\theta)
\end{math}
and
\begin{math}
  \frac{1}{2}\hbar \nabla \cdot ( \rho_{\pm} \vec{u} ) 
  = - \kappa  \rho_{\pm} \pm  \Oxy \sqrt{\rho_+\rho_-} \sin(\theta).
\end{math}
These equations are solved by $\sin(\theta)=0, \rho_+\propto\rho_-$
and thus $\frac{1}{2}m|\vec{u}|^2=\mu+ (\Oxy^2 + \Oz^2)^{1/2}$.  When
desynchronized, a general solution is more complicated, as $\theta$
and $\rho_+/\rho_-$ are time dependent, and one must find the average
$\langle |\vec{u}_{\pm}|^2 \rangle$ in terms of $\langle \mu_{\pm}
\rangle$ .  However, if $\rho_- \gg \rho_+$ in the desynchronized
region, then the real part of Eq.~(\ref{eq:1}) for the majority
component becomes $\frac{1}{2} m \langle |\vec{u}_-|^2 \rangle =
\langle \mu_- \rangle + \Oz$.  The hexagonal vortex lattice resulting
from the interference of three such currents has a spacing $l = (2
/3\sqrt{3})(h/m|\vec{u}|)$.  For the parameters of
Fig.~\ref{fig:vortices}, this gives $l = 1.4 \ell_0$, consistent with
the observed lattice.  For larger $\Oz, \Oxy$, the lattice spacing is
found to vary considerably with $\Oz$; we have checked that the form
discussed here matches the numerical simulations well.

\begin{figure}[htpb]
  \centering
  \includegraphics[width=3.2in]{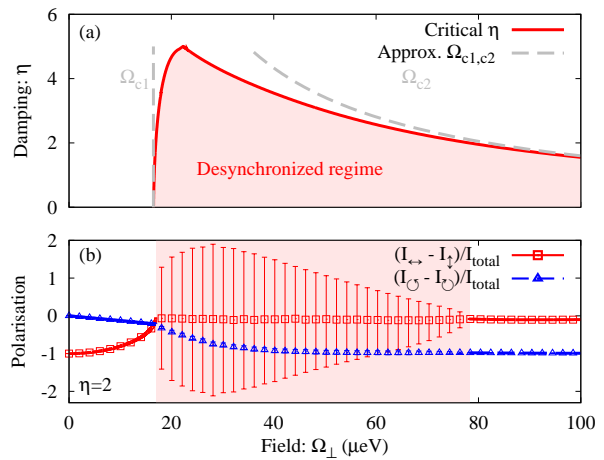}
  \caption{Panel (a): boundary of desynchronized region as a function
    of field $\Oz$ and relaxation rate $\eta$ for homogeneous pumping.
    Solid lines are exact numerical results; gray dashed lines are the
    approximate asymptotes given in the text.  Panel (b): polarization
    degree vs.\ field $\Oz$ for the homogeneous system at $\eta=2$.
    Lines indicate the steady state polarization for the synchronized
    regimes.  The points and error bars indicate the time-average, and
    envelope of polarization oscillations.  (NB the envelopes are
    normalized by the average, rather than instantaneous, intensity.)
  }
  \label{fig:homogeneous}
\end{figure}

We next discuss the critical fields required to switch between
synchronized and desynchronized lattices, as shown in
Fig.~\ref{fig:homogeneous}(a).  
The line in Fig.~\ref{fig:homogeneous} is
calculated for homogeneous pumping; numerical simulations
with inhomogeneous pumping also respect this boundary, i.e. the critical
$\Oz$ for desynchronization is rather robust to inhomogeneity.
Analytical estimates of the critical
fields $\Oz,\Oxy$ help to reveal how they depend on the
system parameters.  The first transition on increasing $\Oz$ is from
the elliptically polarized (synchronized) phase existing for small
$\Oz$ to a desynchronized state.  This transition was discussed in
Refs.~\cite{wouters08:prb,borgh10}.  If $\Oz,\Oxy \ll 2U_1
(\rho_+ +\rho_-)$ (which is the case for the current parameters), then
one may approximate the coupled equations for $\theta,\rho_\pm$ (with
the parametrization $\psi_{\pm} = \sqrt{\rho_{\pm}} e^{i(\phi \pm \theta/2)}$
introduced above) by using standard results for Josephson oscillations
in the Josephson regime~\cite{leggett01}.  The transition thus occurs
at $\Oz = \Omega_{c1} \simeq 2 U_1 \Oxy/ (\Gamma_s - \Gamma_x)$.

If $\eta\neq 0$ there is  another transition at $\Oz=\Omega_{c2}$
back to a synchronized state.  The critical field $\Omega_{c2}$ can be
estimated analytically by linear stability analysis of the
synchronized state at $\Oz>\Omega_{c2}$.  Since $\rho_- \gg \rho_+$ in
this large $\Oz$ limit, one may show that fluctuations of
$\rho_{+}-\rho_{-}$ relax quickly. The remaining part of the
linearized Eq.~(\ref{eq:1}) can be written in terms of
$\zeta=\psi^\ast_+\psi_- $ as:
\begin{equation}
\label{eq:3}
  (1+\eta^2) \hbar \partial_t  \zeta
  =
  (\gamma_{\gss} -i \nu_{\gss}) \zeta
  -
  \Oxy \rho_-
  (\eta + i).
\end{equation}
where $\gamma_{\gss} = 2(\gamma-\kappa) -
\rho_-[\Gamma_s + \Gamma_x + 2\eta(U_0-U_1)]$ and
$\nu_{\text{eff}} = 2 \Oz + [\eta(\Gamma_s - \Gamma_x) -2 U_1] \rho_-$.
Instability of the synchronized state occurs when $\gamma_{\gss}$
becomes negative, thus  to find $\Omega_{c2}$
one thus needs $\rho_-(\Oz)$ at large $\Oz$.  Since $\rho_- \gg
\rho_+$, one may simplify Eq.~(\ref{eq:1}) for $\psi_-$ considerably
to give $\rho_- \simeq (\gamma-\kappa+\eta \Oz)/(\eta U_0 + \Gamma_s)$
hence one has
\begin{equation}
  \label{eq:4}
  \Omega_{c2} \simeq 
  \frac{(\gamma-\kappa)}{\eta}
  \frac{\Gamma_s-\Gamma_x+ 2 \eta U_1}{\Gamma_s+\Gamma_x + 2 \eta(U_0-U_1)}.
\end{equation}
As the relaxation rate $\eta$ increases $\Omega_{c2}$ becomes smaller,
and eventually the desynchronized regime will vanish.

Let us now briefly consider the relation of the experimental results
of \citet{Larionov10} to the polarization desynchronization
transition.  In their experiment, a transition was seen such that for
small $\Oz$ a single frequency elliptically polarized condensate was
observed, and above a critical value of $\Oz$, a transition to a
Zeeman split condensate, with separate frequencies for left- and
right-circular components was observed.  While it was noted
in~\cite{Larionov10} that some features of the experimental results
clearly required aspects of non-equilibrium physics, the transition
was interpreted as a signature of the equilibrium transition to a
circular condensate~\cite{rubo06,*keeling08:spin}.  However, a
fragmented condensate (i.e. having multiple condensate frequencies) is
not expected in equilibrium, but is precisely the signature expected
for the desynchronization transition.  While this identification would
require further experimental corroboration, we have chosen our value
of $\Oxy=3\mu$eV so the desynchronization transition matches this
experimental transition.  Figure~\ref{fig:homogeneous}(b) illustrates
the behavior in the desynchronized regime, showing the amplitude of
polarization fluctuations. The time averaged polarization degree shown
in that figure evolves smoothly with field, and so provides a poor
distinction between synchronized and desynchronized behavior.  One may
however note that both $\Omega_{c2}$, and the equilibrium critical
field~\cite{rubo06,*keeling08:spin} increase with density.  In
contrast, $\Omega_{c1}$, where splitting of the condensate frequencies
should occur due to desynchronization is independent of density (for
small enough $\eta$).  This may provide a clear way to distinguish the
nature of the transition observed experimentally~\cite{Larionov10}.

In conclusion, we have explored the desynchronization transition in a
non-equilibrium spinor polariton condensate, and shown how this
transition can be used to drive a system between lattices of integer
and half vortices.  Using the parameters of recent
experiments\cite{Larionov10}, we show that the conditions for the
desynchronization transition seem likely to be achievable, and note
that aspects of the experimentally observed behavior may even suggest
that the transition already seen is that to a desynchronized state, rather
than to a circularly polarized state.  When the ability to control
polarization via magnetic field is combined with the possibility to
form static vortex lattices via symmetric pumping spots, our proposed
scheme illustrates how one may engineer and control exotic lattice
spin textures deterministically using incoherently pumped polariton
condensates.

\acknowledgments{J.K. acknowledges funding from EPSRC grant
  EP/G004714/2. N.G.B acknowledges EU FP7 ITN project CLERMONT4.}


\begin{thebibliography}{31}%
\makeatletter
\providecommand \@ifxundefined [1]{%
 \@ifx{#1\undefined}
}%
\providecommand \@ifnum [1]{%
 \ifnum #1\expandafter \@firstoftwo
 \else \expandafter \@secondoftwo
 \fi
}%
\providecommand \@ifx [1]{%
 \ifx #1\expandafter \@firstoftwo
 \else \expandafter \@secondoftwo
 \fi
}%
\providecommand \natexlab [1]{#1}%
\providecommand \enquote  [1]{``#1''}%
\providecommand \bibnamefont  [1]{#1}%
\providecommand \bibfnamefont [1]{#1}%
\providecommand \citenamefont [1]{#1}%
\providecommand \href@noop [0]{\@secondoftwo}%
\providecommand \href [0]{\begingroup \@sanitize@url \@href}%
\providecommand \@href[1]{\@@startlink{#1}\@@href}%
\providecommand \@@href[1]{\endgroup#1\@@endlink}%
\providecommand \@sanitize@url [0]{\catcode `\\12\catcode `\$12\catcode
  `\&12\catcode `\#12\catcode `\^12\catcode `\_12\catcode `\%12\relax}%
\providecommand \@@startlink[1]{}%
\providecommand \@@endlink[0]{}%
\providecommand \url  [0]{\begingroup\@sanitize@url \@url }%
\providecommand \@url [1]{\endgroup\@href {#1}{\urlprefix }}%
\providecommand \urlprefix  [0]{URL }%
\providecommand \Eprint [0]{\href }%
\providecommand \doibase [0]{http://dx.doi.org/}%
\providecommand \selectlanguage [0]{\@gobble}%
\providecommand \bibinfo  [0]{\@secondoftwo}%
\providecommand \bibfield  [0]{\@secondoftwo}%
\providecommand \translation [1]{[#1]}%
\providecommand \BibitemOpen [0]{}%
\providecommand \bibitemStop [0]{}%
\providecommand \bibitemNoStop [0]{.\EOS\space}%
\providecommand \EOS [0]{\spacefactor3000\relax}%
\providecommand \BibitemShut  [1]{\csname bibitem#1\endcsname}%
\let\auto@bib@innerbib\@empty
\bibitem [{\citenamefont {Volovik}(2003)}]{volovik03}%
  \BibitemOpen
  \bibfield  {author} {\bibinfo {author} {\bibfnamefont {G.~E.}\ \bibnamefont
  {Volovik}},\ }\href@noop {} {\emph {\bibinfo {title} {The Universe in a
  Helium Droplet}}}\ (\bibinfo  {publisher} {Oxford University Press},\
  \bibinfo {address} {Oxford},\ \bibinfo {year} {2003})\BibitemShut {NoStop}%
\bibitem [{\citenamefont {Ruben}\ \emph {et~al.}(2010)\citenamefont {Ruben},
  \citenamefont {Morgan},\ and\ \citenamefont {Paganin}}]{Ruben10}%
  \BibitemOpen
  \bibfield  {author} {\bibinfo {author} {\bibfnamefont {G.}~\bibnamefont
  {Ruben}}, \bibinfo {author} {\bibfnamefont {M.}~\bibnamefont {Morgan}}, \
  and\ \bibinfo {author} {\bibfnamefont {D.}~\bibnamefont {Paganin}},\ }\href
  {\doibase 10.1103/PhysRevLett.105.220402} {\bibfield  {journal} {\bibinfo
  {journal} {Phys. Rev. Lett.}\ }\textbf {\bibinfo {volume} {105}},\ \bibinfo
  {pages} {220402} (\bibinfo {year} {2010})}\BibitemShut {NoStop}%
\bibitem [{\citenamefont {Pismen}(1994)}]{Pismen94}%
  \BibitemOpen
  \bibfield  {author} {\bibinfo {author} {\bibfnamefont {L.}~\bibnamefont
  {Pismen}},\ }\href {\doibase 10.1103/PhysRevLett.72.2557} {\bibfield
  {journal} {\bibinfo  {journal} {Phys. Rev. Lett.}\ }\textbf {\bibinfo
  {volume} {72}},\ \bibinfo {pages} {2557} (\bibinfo {year}
  {1994})}\BibitemShut {NoStop}%
\bibitem [{\citenamefont {Mackenzie}\ and\ \citenamefont
  {Maeno}(2003)}]{mackenzie03}%
  \BibitemOpen
  \bibfield  {author} {\bibinfo {author} {\bibfnamefont {A.~P.}\ \bibnamefont
  {Mackenzie}}\ and\ \bibinfo {author} {\bibfnamefont {Y.}~\bibnamefont
  {Maeno}},\ }\href {\doibase 10.1103/RevModPhys.75.657} {\bibfield  {journal}
  {\bibinfo  {journal} {Rev. Mod. Phys.}\ }\textbf {\bibinfo {volume} {75}},\
  \bibinfo {pages} {657} (\bibinfo {year} {2003})}\BibitemShut {NoStop}%
\bibitem [{\citenamefont {Deng}\ \emph {et~al.}(2010)\citenamefont {Deng},
  \citenamefont {Haug},\ and\ \citenamefont {Yamamoto}}]{deng10}%
  \BibitemOpen
  \bibfield  {author} {\bibinfo {author} {\bibfnamefont {H.}~\bibnamefont
  {Deng}}, \bibinfo {author} {\bibfnamefont {H.}~\bibnamefont {Haug}}, \ and\
  \bibinfo {author} {\bibfnamefont {Y.}~\bibnamefont {Yamamoto}},\ }\href
  {\doibase 10.1103/RevModPhys.82.1489} {\bibfield  {journal} {\bibinfo
  {journal} {Rev. Mod. Phys.}\ }\textbf {\bibinfo {volume} {82}},\ \bibinfo
  {pages} {1489} (\bibinfo {year} {2010})}\BibitemShut {NoStop}%
\bibitem [{\citenamefont {Stali{\=u}nas}\ and\ \citenamefont
  {S{\'a}nchez-Morcillo}(2003)}]{staliunas03}%
  \BibitemOpen
  \bibfield  {author} {\bibinfo {author} {\bibfnamefont {K.}~\bibnamefont
  {Stali{\=u}nas}}\ and\ \bibinfo {author} {\bibfnamefont {V.~J.}\ \bibnamefont
  {S{\'a}nchez-Morcillo}},\ }\href@noop {} {\emph {\bibinfo {title} {Transverse
  Patterns in Nonlinear Optical Resonators}}},\ \bibinfo {series} {Springer
  Tracts in Modern Physics}, Vol.\ \bibinfo {volume} {183}\ (\bibinfo
  {publisher} {Springer-Verlag},\ \bibinfo {address} {Berlin},\ \bibinfo {year}
  {2003})\BibitemShut {NoStop}%
\bibitem [{\citenamefont {Aranson}\ and\ \citenamefont
  {Kramer}(2002)}]{aranson02}%
  \BibitemOpen
  \bibfield  {author} {\bibinfo {author} {\bibfnamefont {I.~S.}\ \bibnamefont
  {Aranson}}\ and\ \bibinfo {author} {\bibfnamefont {L.}~\bibnamefont
  {Kramer}},\ }\href {\doibase 10.1103/RevModPhys.74.99} {\bibfield  {journal}
  {\bibinfo  {journal} {Rev. Mod. Phys.}\ }\textbf {\bibinfo {volume} {74}},\
  \bibinfo {pages} {99} (\bibinfo {year} {2002})}\BibitemShut {NoStop}%
\bibitem [{\citenamefont {Lagoudakis}\ \emph {et~al.}(2008)\citenamefont
  {Lagoudakis} \emph {et~al.}}]{lagoudakis08}%
  \BibitemOpen
  \bibfield  {author} {\bibinfo {author} {\bibfnamefont {K.~G.}\ \bibnamefont
  {Lagoudakis}} \emph {et~al.},\ }\href {\doibase 10.1038/nphys1051} {\bibfield
   {journal} {\bibinfo  {journal} {Nature Phys.}\ }\textbf {\bibinfo {volume}
  {4}},\ \bibinfo {pages} {706} (\bibinfo {year} {2008})}\BibitemShut {NoStop}%
\bibitem [{\citenamefont {Amo}\ \emph {et~al.}(2009)\citenamefont {Amo} \emph
  {et~al.}}]{amo09:bullet}%
  \BibitemOpen
  \bibfield  {author} {\bibinfo {author} {\bibfnamefont {A.}~\bibnamefont
  {Amo}} \emph {et~al.},\ }\href {\doibase 10.1038/nature07640} {\bibfield
  {journal} {\bibinfo  {journal} {Nature}\ }\textbf {\bibinfo {volume} {457}},\
  \bibinfo {pages} {291} (\bibinfo {year} {2009})}\BibitemShut {NoStop}%
\bibitem [{\citenamefont {Wertz}\ \emph {et~al.}(2010)\citenamefont {Wertz}
  \emph {et~al.}}]{wertz10}%
  \BibitemOpen
  \bibfield  {author} {\bibinfo {author} {\bibfnamefont {E.}~\bibnamefont
  {Wertz}} \emph {et~al.},\ }\href {\doibase 10.1038/nphys1750} {\bibfield
  {journal} {\bibinfo  {journal} {Nat. Phys.}\ }\textbf {\bibinfo {volume}
  {6}},\ \bibinfo {pages} {860} (\bibinfo {year} {2010})}\BibitemShut {NoStop}%
\bibitem [{\citenamefont {Rubo}(2007)}]{rubo07}%
  \BibitemOpen
  \bibfield  {author} {\bibinfo {author} {\bibfnamefont {Y.~G.}\ \bibnamefont
  {Rubo}},\ }\href {\doibase 10.1103/PhysRevLett.99.106401} {\bibfield
  {journal} {\bibinfo  {journal} {Phys. Rev. Lett.}\ }\textbf {\bibinfo
  {volume} {99}},\ \bibinfo {pages} {106401} (\bibinfo {year}
  {2007})}\BibitemShut {NoStop}%
\bibitem [{\citenamefont {Lagoudakis}\ \emph {et~al.}(2009)\citenamefont
  {Lagoudakis} \emph {et~al.}}]{lagoudakis09}%
  \BibitemOpen
  \bibfield  {author} {\bibinfo {author} {\bibfnamefont {K.~G.}\ \bibnamefont
  {Lagoudakis}} \emph {et~al.},\ }\href {\doibase 10.1126/science.1177980}
  {\bibfield  {journal} {\bibinfo  {journal} {Science}\ }\textbf {\bibinfo
  {volume} {326}},\ \bibinfo {pages} {974} (\bibinfo {year}
  {2009})}\BibitemShut {NoStop}%
\bibitem [{\citenamefont {Flayac}\ \emph {et~al.}(2010)\citenamefont {Flayac},
  \citenamefont {Shelykh}, \citenamefont {Solnyshkov},\ and\ \citenamefont
  {Malpuech}}]{Flayac10}%
  \BibitemOpen
  \bibfield  {author} {\bibinfo {author} {\bibfnamefont {H.}~\bibnamefont
  {Flayac}}, \bibinfo {author} {\bibfnamefont {I.~A.}\ \bibnamefont {Shelykh}},
  \bibinfo {author} {\bibfnamefont {D.~D.}\ \bibnamefont {Solnyshkov}}, \ and\
  \bibinfo {author} {\bibfnamefont {G.}~\bibnamefont {Malpuech}},\ }\href
  {\doibase 10.1103/PhysRevB.81.045318} {\bibfield  {journal} {\bibinfo
  {journal} {Phys. Rev. B}\ }\textbf {\bibinfo {volume} {81}},\ \bibinfo
  {pages} {045318} (\bibinfo {year} {2010})}\BibitemShut {NoStop}%
\bibitem [{\citenamefont {Solano}\ and\ \citenamefont
  {Rubo}(2010)}]{solano10-2}%
  \BibitemOpen
  \bibfield  {author} {\bibinfo {author} {\bibfnamefont {M.~T.}\ \bibnamefont
  {Solano}}\ and\ \bibinfo {author} {\bibfnamefont {Y.~G.}\ \bibnamefont
  {Rubo}},\ }\href {\doibase 10.1103/PhysRevB.82.127301} {\bibfield  {journal}
  {\bibinfo  {journal} {Phys. Rev. B}\ }\textbf {\bibinfo {volume} {82}},\
  \bibinfo {pages} {127301} (\bibinfo {year} {2010})}\BibitemShut {NoStop}%
\bibitem [{\citenamefont {Keeling}\ and\ \citenamefont
  {Berloff}(2008)}]{keeling08:gpe}%
  \BibitemOpen
  \bibfield  {author} {\bibinfo {author} {\bibfnamefont {J.}~\bibnamefont
  {Keeling}}\ and\ \bibinfo {author} {\bibfnamefont {N.~G.}\ \bibnamefont
  {Berloff}},\ }\href {\doibase 10.1103/PhysRevLett.100.250401} {\bibfield
  {journal} {\bibinfo  {journal} {Phys. Rev. Lett.}\ }\textbf {\bibinfo
  {volume} {100}},\ \bibinfo {pages} {250401} (\bibinfo {year}
  {2008})}\BibitemShut {NoStop}%
\bibitem [{\citenamefont {Roumpos}\ \emph {et~al.}(2011)\citenamefont {Roumpos}
  \emph {et~al.}}]{roumpos10}%
  \BibitemOpen
  \bibfield  {author} {\bibinfo {author} {\bibfnamefont {G.}~\bibnamefont
  {Roumpos}} \emph {et~al.},\ }\href {\doibase 10.1038/nphys1841} {\bibfield
  {journal} {\bibinfo  {journal} {Nat. Phys.}\ }\textbf {\bibinfo {volume}
  {7}},\ \bibinfo {pages} {129} (\bibinfo {year} {2011})}\BibitemShut {NoStop}%
\bibitem [{\citenamefont {Lagoudakis}\ \emph {et~al.}(2011)\citenamefont
  {Lagoudakis} \emph {et~al.}}]{lagoudakis11}%
  \BibitemOpen
  \bibfield  {author} {\bibinfo {author} {\bibfnamefont {K.~G.}\ \bibnamefont
  {Lagoudakis}} \emph {et~al.},\ }\href {\doibase
  10.1103/PhysRevLett.106.115301} {\bibfield  {journal} {\bibinfo  {journal}
  {Phys. Rev. Lett.}\ }\textbf {\bibinfo {volume} {106}},\ \bibinfo {pages}
  {115301} (\bibinfo {year} {2011})}\BibitemShut {NoStop}%
\bibitem [{\citenamefont {Gorbach}\ \emph {et~al.}(2010)\citenamefont
  {Gorbach}, \citenamefont {Hartley},\ and\ \citenamefont
  {Skryabin}}]{gorbach10}%
  \BibitemOpen
  \bibfield  {author} {\bibinfo {author} {\bibfnamefont {A.~V.}\ \bibnamefont
  {Gorbach}}, \bibinfo {author} {\bibfnamefont {R.}~\bibnamefont {Hartley}}, \
  and\ \bibinfo {author} {\bibfnamefont {D.~V.}\ \bibnamefont {Skryabin}},\
  }\href {\doibase 10.1103/PhysRevLett.104.213903} {\bibfield  {journal}
  {\bibinfo  {journal} {Phys. Rev. Lett.}\ }\textbf {\bibinfo {volume} {104}},\
  \bibinfo {pages} {213903} (\bibinfo {year} {2010})}\BibitemShut {NoStop}%
\bibitem [{\citenamefont {Pigeon}\ \emph {et~al.}(2011)\citenamefont {Pigeon},
  \citenamefont {Carusotto},\ and\ \citenamefont {Ciuti}}]{pigeon11}%
  \BibitemOpen
  \bibfield  {author} {\bibinfo {author} {\bibfnamefont {S.}~\bibnamefont
  {Pigeon}}, \bibinfo {author} {\bibfnamefont {I.}~\bibnamefont {Carusotto}}, \
  and\ \bibinfo {author} {\bibfnamefont {C.}~\bibnamefont {Ciuti}},\ }\href
  {\doibase 10.1103/PhysRevB.83.144513} {\bibfield  {journal} {\bibinfo
  {journal} {Phys. Rev. B}\ }\textbf {\bibinfo {volume} {83}},\ \bibinfo
  {pages} {144513} (\bibinfo {year} {2011})}\BibitemShut {NoStop}%
\bibitem [{\citenamefont {Liew}\ \emph {et~al.}(2008)\citenamefont {Liew},
  \citenamefont {Rubo},\ and\ \citenamefont {Kavokin}}]{Liew08}%
  \BibitemOpen
  \bibfield  {author} {\bibinfo {author} {\bibfnamefont {T.~C.~H.}\
  \bibnamefont {Liew}}, \bibinfo {author} {\bibfnamefont {Y.~G.}\ \bibnamefont
  {Rubo}}, \ and\ \bibinfo {author} {\bibfnamefont {A.}~\bibnamefont
  {Kavokin}},\ }\href {\doibase 10.1103/PhysRevLett.101.187401} {\bibfield
  {journal} {\bibinfo  {journal} {Phys. Rev. Lett.}\ }\textbf {\bibinfo
  {volume} {101}},\ \bibinfo {pages} {187401} (\bibinfo {year}
  {2008})}\BibitemShut {NoStop}%
\bibitem [{\citenamefont {Wouters}(2008)}]{wouters08:prb}%
  \BibitemOpen
  \bibfield  {author} {\bibinfo {author} {\bibfnamefont {M.}~\bibnamefont
  {Wouters}},\ }\href {\doibase 10.1103/PhysRevB.77.121302} {\bibfield
  {journal} {\bibinfo  {journal} {Phys. Rev. B}\ }\textbf {\bibinfo {volume}
  {77}},\ \bibinfo {pages} {121302} (\bibinfo {year} {2008})}\BibitemShut
  {NoStop}%
\bibitem [{\citenamefont {Borgh}\ \emph {et~al.}(2010)\citenamefont {Borgh},
  \citenamefont {Keeling},\ and\ \citenamefont {Berloff}}]{borgh10}%
  \BibitemOpen
  \bibfield  {author} {\bibinfo {author} {\bibfnamefont {M.~O.}\ \bibnamefont
  {Borgh}}, \bibinfo {author} {\bibfnamefont {J.}~\bibnamefont {Keeling}}, \
  and\ \bibinfo {author} {\bibfnamefont {N.~G.}\ \bibnamefont {Berloff}},\
  }\href {\doibase 10.1103/PhysRevB.81.235302} {\bibfield  {journal} {\bibinfo
  {journal} {Phys. Rev. B}\ }\textbf {\bibinfo {volume} {81}},\ \bibinfo
  {pages} {235302} (\bibinfo {year} {2010})}\BibitemShut {NoStop}%
\bibitem [{\citenamefont {Wouters}\ and\ \citenamefont
  {Carusotto}(2010)}]{wouters10:superfluid}%
  \BibitemOpen
  \bibfield  {author} {\bibinfo {author} {\bibfnamefont {M.}~\bibnamefont
  {Wouters}}\ and\ \bibinfo {author} {\bibfnamefont {I.}~\bibnamefont
  {Carusotto}},\ }\href {\doibase 10.1103/PhysRevLett.105.020602} {\bibfield
  {journal} {\bibinfo  {journal} {Phys. Rev. Lett.}\ }\textbf {\bibinfo
  {volume} {105}},\ \bibinfo {pages} {020602} (\bibinfo {year}
  {2010})}\BibitemShut {NoStop}%
\bibitem [{\citenamefont {Wouters}\ and\ \citenamefont {Savona}()}]{wouters10}%
  \BibitemOpen
  \bibfield  {author} {\bibinfo {author} {\bibfnamefont {M.}~\bibnamefont
  {Wouters}}\ and\ \bibinfo {author} {\bibfnamefont {V.}~\bibnamefont
  {Savona}},\ }\href@noop {} {}\Eprint {http://arxiv.org/abs/arXiv:1007.5453}
  {arXiv:1007.5453} \BibitemShut {NoStop}%
\bibitem [{\citenamefont {Wouters}\ \emph {et~al.}(2010)\citenamefont
  {Wouters}, \citenamefont {Liew},\ and\ \citenamefont {Savona}}]{liew10}%
  \BibitemOpen
  \bibfield  {author} {\bibinfo {author} {\bibfnamefont {M.}~\bibnamefont
  {Wouters}}, \bibinfo {author} {\bibfnamefont {T.~C.~H.}\ \bibnamefont
  {Liew}}, \ and\ \bibinfo {author} {\bibfnamefont {V.}~\bibnamefont
  {Savona}},\ }\href {\doibase 10.1103/PhysRevB.82.245315} {\bibfield
  {journal} {\bibinfo  {journal} {Phys. Rev. B}\ }\textbf {\bibinfo {volume}
  {82}},\ \bibinfo {pages} {245315} (\bibinfo {year} {2010})}\BibitemShut
  {NoStop}%
\bibitem [{\citenamefont {Griffin}\ \emph {et~al.}(2009)\citenamefont
  {Griffin}, \citenamefont {Nikuni},\ and\ \citenamefont
  {Zaremba}}]{griffin09}%
  \BibitemOpen
  \bibfield  {author} {\bibinfo {author} {\bibfnamefont {A.}~\bibnamefont
  {Griffin}}, \bibinfo {author} {\bibfnamefont {T.}~\bibnamefont {Nikuni}}, \
  and\ \bibinfo {author} {\bibfnamefont {E.}~\bibnamefont {Zaremba}},\
  }\href@noop {} {\emph {\bibinfo {title} {Bose-Condensed Gases at Finite
  Temperatures}}}\ (\bibinfo  {publisher} {Cambridge University Press},\
  \bibinfo {address} {Cambridge},\ \bibinfo {year} {2009})\BibitemShut
  {NoStop}%
\bibitem [{\citenamefont {Larionov}\ \emph {et~al.}(2010)\citenamefont
  {Larionov} \emph {et~al.}}]{Larionov10}%
  \BibitemOpen
  \bibfield  {author} {\bibinfo {author} {\bibfnamefont {A.}~\bibnamefont
  {Larionov}} \emph {et~al.},\ }\href {\doibase 10.1103/PhysRevLett.105.256401}
  {\bibfield  {journal} {\bibinfo  {journal} {Phys. Rev. Lett.}\ }\textbf
  {\bibinfo {volume} {105}},\ \bibinfo {pages} {256401} (\bibinfo {year}
  {2010})}\BibitemShut {NoStop}%
\bibitem [{\citenamefont {Berloff}()}]{turbulence10}%
  \BibitemOpen
  \bibfield  {author} {\bibinfo {author} {\bibfnamefont {N.~G.}\ \bibnamefont
  {Berloff}},\ }\href@noop {} {}\Eprint {http://arxiv.org/abs/arXiv:1010.5225}
  {arXiv:1010.5225} \BibitemShut {NoStop}%
\bibitem [{\citenamefont {Leggett}(2001)}]{leggett01}%
  \BibitemOpen
  \bibfield  {author} {\bibinfo {author} {\bibfnamefont {A.~J.}\ \bibnamefont
  {Leggett}},\ }\href@noop {} {\bibfield  {journal} {\bibinfo  {journal} {Rev.
  Mod. Phys.}\ }\textbf {\bibinfo {volume} {73}},\ \bibinfo {pages} {307}
  (\bibinfo {year} {2001})}\BibitemShut {NoStop}%
\bibitem [{\citenamefont {Rubo}\ \emph {et~al.}(2006)\citenamefont {Rubo},
  \citenamefont {Kavokin},\ and\ \citenamefont {Shelykh}}]{rubo06}%
  \BibitemOpen
  \bibfield  {author} {\bibinfo {author} {\bibfnamefont {Y.~G.}\ \bibnamefont
  {Rubo}}, \bibinfo {author} {\bibfnamefont {A.~V.}\ \bibnamefont {Kavokin}}, \
  and\ \bibinfo {author} {\bibfnamefont {I.~A.}\ \bibnamefont {Shelykh}},\
  }\href {\doibase 10.1016/j.physleta.2006.05.015} {\bibfield  {journal}
  {\bibinfo  {journal} {Phys. Lett. A}\ }\textbf {\bibinfo {volume} {358}},\
  \bibinfo {pages} {227} (\bibinfo {year} {2006})}\BibitemShut {NoStop}%
\bibitem [{\citenamefont {Keeling}(2008)}]{keeling08:spin}%
  \BibitemOpen
  \bibfield  {author} {\bibinfo {author} {\bibfnamefont {J.}~\bibnamefont
  {Keeling}},\ }\href {\doibase 10.1103/PhysRevB.78.205316} {\bibfield
  {journal} {\bibinfo  {journal} {Phys. Rev. B}\ }\textbf {\bibinfo {volume}
  {78}},\ \bibinfo {pages} {205316} (\bibinfo {year} {2008})}\BibitemShut
  {NoStop}%
\end{thebibliography}
%

\end{document}